\documentclass[12pt,a4paper]{article}
\makeatletter
\def\eqnarray{%
\stepcounter{equation}%
\let\@currentlabel=\theequation
\global\@eqnswtrue
\global\@eqcnt\z@
\tabskip\@centering
\let\\=\@eqncr
$$\halign to \displaywidth\bgroup\@eqnsel\hskip\@centering
$\displaystyle\tabskip\z@{##}$&\global\@eqcnt\@ne
\hfil$\displaystyle{{}##{}}$\hfil
&\global\@eqcnt\tw@$\displaystyle\tabskip\z@{##}$\hfil
\tabskip\@centering&\llap{##}\tabskip\z@\cr}
\makeatother
\newcommand{\ket}[1]{{\vert{#1}\rangle}}
\newcommand{\bra}[1]{{\langle{#1}\vert}}
\newcommand{\kett}[1]{{\vert{#1}\rangle\rangle}}
\newcommand{\braa}[1]{{\langle\langle{#1}\vert}}
\newcommand{\calh}{{\cal H}}

\newcommand{\fukuso}{{\mathbf C}}
\newcommand{\seisu}{{\mathbf Z}}
\newcommand{\futon}{{\bf N}}

\begin{document}

\title{\sl Cavity QED and Quantum Computation in the Strong Coupling Regime}
\author{
  Kazuyuki FUJII
  \thanks{E-mail address : fujii@yokohama-cu.ac.jp }\\
  Department of Mathematical Sciences\\
  Yokohama City University\\
  Yokohama, 236-0027\\
  Japan
  }
\date{}
\maketitle
%
%
%
%
\begin{abstract}
 In this paper we propose a Hamiltonian generalizing the interaction 
 of the two--level atom and both the single radiation mode and external field 
 $\cdots$ a kind of cavity QED. 
 We solve the Schrodinger equation in the strong coupling regime by making 
 use of rotating wave approximation under new resonance conditions containing 
 the Bessel functions and etc, and 
 obtain unitary transformations of four types corresponding to Rabi 
 oscillations which perform quantum logic gates in Quantum Computation. 
\end{abstract}


%
%
%
%
In this paper we consider a unified model of the interaction of the two--level 
atom and both the single radiation mode and external field (periodic usually)
in a cavity. We deal with the external field as a classical one. 
As a general introduction to this topic in Quantum Optics see \cite{AE}, 
\cite{MSIII}, \cite{C-HDG}. 
Our model is deeply related to the one of trapped ions in a cavity with 
the photon interaction (Cavity QED). 

In our model we are especially interested in the strong coupling regime, 
\cite{MFr2}, \cite{KF2}, \cite{KF3}. One of motivations is a recent very 
interesting experiment, \cite{NPT}. See \cite{C-HDG} and \cite{MFr3} 
as a general introduction. 

In \cite{MFr2} and \cite{KF2} we have treated the strong coupling regime of 
the interaction model of the two--level atom and the single radiation mode, 
and have given some explicit solutions under the resonance conditions and 
rotating wave approximations. 

On the other hand we want to add some external field (like Laser one) to 
the above model which will make the model more realistic (for example in 
Quantum Computation). Therefore we propose the unified model. 

We would like to solve our model in the strong coupling regime. 
Especially we want to show the existence of Rabi oscillations in this regime 
because the real purpose of a series of study (\cite{KF2}, \cite{KF3}, 
\cite{KF7}) is an application to Quantum Computation (see \cite{KF1} 
as a brief introduction to it). 

In this paper we solve the Schrodinger equations in this regime by 
making use of rotating wave approximation under new resonance conditions 
containing the Bessel functions and etc, and obtain a unitary transformation 
in one qubit case and unitary transformations of four types in two qubit case, 
which perform quantum logic gates. 

Our solutions might give a new insight into Quantum Optics or Condensed Matter 
Physics as well as Quantum Computation. 

\par \vspace{5mm}
Let $\{\sigma_{1}, \sigma_{2}, \sigma_{3}\}$ be Pauli matrices and 
${\bf 1}_{2}$ a unit matrix : 
\begin{equation}
\sigma_{1} = 
\left(
  \begin{array}{cc}
    0& 1 \\
    1& 0
  \end{array}
\right), \quad 
\sigma_{2} = 
\left(
  \begin{array}{cc}
    0& -i \\
    i& 0
  \end{array}
\right), \quad 
\sigma_{3} = 
\left(
  \begin{array}{cc}
    1& 0 \\
    0& -1
  \end{array}
\right), \quad 
{\bf 1}_{2} = 
\left(
  \begin{array}{cc}
    1& 0 \\
    0& 1
  \end{array}
\right), 
\end{equation}
and 
$\sigma_{+} = (1/2)(\sigma_{1}+i\sigma_{1})$, 
$\sigma_{-} = (1/2)(\sigma_{1}-i\sigma_{1})$. 
Let $W$ be the Walsh--Hadamard matrix 
\begin{equation}
\label{eq:2-Walsh-Hadamard}
W=\frac{1}{\sqrt{2}}
\left(
  \begin{array}{cc}
    1& 1 \\
    1& -1
  \end{array}
\right)
=W^{-1}\ , 
\end{equation}
then we can diagonalize $\sigma_{1}$ as 
$
\sigma_{1}=W\sigma_{3}W^{-1}=W\sigma_{3}W
$ 
by making use of this $W$. 
The eigenvalues of $\sigma_{1}$ is $\{1,-1\}$ with eigenvectors 
\begin{equation}
\label{eq:eigenvectors of sigma}
\ket{1}=\frac{1}{\sqrt{2}}
\left(
  \begin{array}{c}
    1 \\
    1
  \end{array}
\right), \quad 
\ket{-1}=\frac{1}{\sqrt{2}}
\left(
  \begin{array}{c}
    1 \\
    -1
  \end{array}
\right)
\quad \Longrightarrow \quad 
\ket{\lambda}=\frac{1}{\sqrt{2}}
\left(
  \begin{array}{c}
    1 \\
    \lambda
  \end{array}
\right).
\end{equation}

Let us consider an atom with $2$ energy levels $E_{0}$ and $E_{1}$ (of course 
$E_{1} > E_{0}$). 
Its Hamiltonian is in the diagonal form given as 
\begin{equation}
H_{0}=
\left(
  \begin{array}{cc}
    E_{0}& 0 \\
    0& E_{1}
  \end{array}
\right).
\end{equation}
This is rewritten as 
\begin{equation}
\label{eq:2}
H_{0}=
\frac{E_{0}+E_{1}}{2}
\left(
  \begin{array}{cc}
    1& 0 \\
    0& 1
  \end{array}
\right)-
\frac{E_{1}-E_{0}}{2}
\left(
  \begin{array}{cc}
    1& 0 \\
    0& -1
  \end{array}
\right)
\equiv \Delta_{0}{\bf 1}_{2}-\frac{\Delta}{2}\sigma_{3},
\end{equation}
where $\Delta=E_{1}-E_{0}$ is a energy difference. 
Since we usually take no interest in constant terms, we set 
\begin{equation}
\label{eq:2-energy-hamiltonian}
H_{0}=-\frac{\Delta}{2}\sigma_{3}. 
\end{equation}

We consider an atom with two energy levels which interacts with external 
(periodic) field with $g\mbox{cos}(\omega_{E}t)$. 
In the following we set $\hbar=1$ for simplicity. 
The Hamiltonian in the dipole approximation is given by 
\begin{equation}
\label{eq:2-hamiltonian}
H=H_{0}+g \mbox{cos}(\omega_{E}t+\phi)\sigma_{1}
=-\frac{\Delta}{2}\sigma_{3}+g\ \mbox{cos}(\omega_{E}t+\phi)\sigma_{1}, 
\end{equation}
where $\omega_{E}$ is the frequency of the external field, $g$ the coupling 
constant between the external field and the atom. 
We note that to solve this model without assuming the rotating wave 
approximation is not easy, see \cite{MFr1}, \cite{MFr4}, 
\cite{BaWr}, \cite{SGD}, \cite{CEC}. 

\par \noindent 
In the following we change the sign in the kinetic term, namely from 
$-\Delta/2$ to $\Delta/2$, to set the model for other models. 
However this is minor.

\vspace{5mm} \par 
Now we make a short review of the harmonic oscillator within our necessity.
Let $a(a^\dagger)$ be the annihilation (creation) operator of the harmonic 
oscillator.
If we set $N\equiv a^\dagger a$ (:\ number operator), then we have 
\begin{equation}
  \label{eq:basic-1}
  [N,a^\dagger]=a^\dagger\ ,\
  [N,a]=-a\ ,\
  [a^\dagger, a]=-\mathbf{1}\ .
\end{equation}
Let $\calh$ be a Fock space generated by $a$ and $a^\dagger$, and
$\{\ket{n}\vert\  n\in\futon\cup\{0\}\}$ be its basis.
The actions of $a$ and $a^\dagger$ on $\calh$ are given by
\begin{equation}
  \label{eq:basic-2}
  a\ket{n} = \sqrt{n}\ket{n-1}\ ,\
  a^{\dagger}\ket{n} = \sqrt{n+1}\ket{n+1}\ ,
  N\ket{n} = n\ket{n}
\end{equation}
where $\ket{0}$ is a normalized vacuum ($a\ket{0}=0\  {\rm and}\  
\langle{0}\vert{0}\rangle = 1$). From (\ref{eq:basic-2})
state $\ket{n}$ for $n \geq 1$ are given by
\begin{equation}
  \label{eq:basic-3}
  \ket{n} = \frac{(a^{\dagger})^{n}}{\sqrt{n!}}\ket{0}\ .
\end{equation}
These states satisfy the orthogonality and completeness conditions 
\begin{equation}
  \label{eq:basic-4}
   \langle{m}\vert{n}\rangle = \delta_{mn}\ ,\quad \sum_{n=0}^{\infty}
   \ket{n}\bra{n} = \mathbf{1}\ . 
\end{equation}
Then the displacement (coherent) operator and coherent state are defined as
\begin{equation}
  \label{eq:basic-5}
      D(z) = \mbox{e}^{za^{\dagger}- \bar{z}a}\ ;\quad 
      \ket{z} = D(z)\ket{0}\quad \mbox{for} \quad z \in \fukuso .  
\end{equation}

We consider the quantum theory of the interaction between an atom with 
two--energy levels and single radiation mode (a harmonic oscillator). 
The Hamiltonian in this case is 
\begin{equation}
\label{eq:hamiltonian-(0)}
H=\omega {\bf 1}_{2}\otimes a^{\dagger}a + 
\frac{\Delta}{2}\sigma_{3}\otimes {\bf 1} +
g\sigma_{1}\otimes (a^{\dagger}+a) 
\end{equation}
where $\omega$ is the frequency of the radiation mode, $g$ the coupling 
between the radiation field and the atom, see for example \cite{C-HDG}, 
\cite{MFr2}. 

Now it is very natural for us to include (\ref{eq:2-hamiltonian}) into 
(\ref{eq:hamiltonian-(0)}), so we present the following 

\vspace{3mm} \par \noindent
{\bf General Hamiltonian}
\begin{equation}
\label{eq:hamiltonian-(i)}
H=\omega{\bf 1}_{2}\otimes a^{\dagger}a + 
g_{1}\sigma_{1}\otimes (a^{\dagger}+a) + 
\frac{\Delta}{2}\sigma_{3}\otimes {\bf 1} + 
g_{2} \mbox{cos}(\omega_{E}t+\phi)\sigma_{1}\otimes {\bf 1}. 
\end{equation}
Our Hamiltonian has two coupling constants.\ 
We note that our model is deeply related to the Cavity QED 
(trapped ions in a cavity with the photon interaction). 

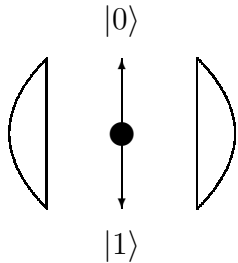
\begin{center}
\setlength{\unitlength}{1mm} 
\begin{picture}(80,40)(0,-10)
\bezier{200}(20,0)(10,10)(20,20)
\put(20,0){\line(0,1){20}}
\put(20,20){\makebox(20,10)[c]{$|0\rangle$}}
\put(30,10){\vector(0,1){10}}
\put(30,10){\vector(0,-1){10}}
\put(20,-10){\makebox(20,10)[c]{$|1\rangle$}}
\put(30,10){\circle*{3}}
\bezier{200}(40,0)(50,10)(40,20)
\put(40,0){\line(0,1){20}}
\end{picture}
\end{center}
This Hamiltonian is also related to the one presented recently by 
Sch{\"o}n and Cirac \cite{SCi}
\vspace{2mm}\par \noindent
\begin{eqnarray}
\label{eq:hamiltonian-(ii)}
H=\frac{p^2}{2m}&+& 
\omega_{0}{\bf 1}_{2}\otimes a^{\dagger}a +
g(x)\left(\sigma_{+}\otimes a+\sigma_{-}\otimes a^{\dagger}\right) +
\nonumber \\
&&\frac{\omega_0}{2}\sigma_{3}\otimes {\bf 1}+ 
\frac{\Omega}{2}
\left(
\mbox{e}^{-i\omega_{L}t}\ \sigma_{+}\otimes {\bf 1} + 
\mbox{e}^{i\omega_{L}t}\ \sigma_{-}\otimes {\bf 1}
\right). 
\end{eqnarray}
For the meaning of several constants see \cite{SCi}. 
They have assumed the rotating wave approximation (see for example 
\cite{C-HDG}) and the resonance condition, and 
use a position--dependent coupling constant $g(x)$, 
so their model is different from ours in these points.  

A comment is in order. Following \cite{SCi} 
the Hamiltonian (\ref{eq:hamiltonian-(i)}) might be modified to 
\begin{equation}
H=\frac{p^2}{2m} + \omega{\bf 1}_{2}\otimes a^{\dagger}a + 
g_{1}(x)\sigma_{1}\otimes (a^{\dagger}+a) + 
\frac{\Delta}{2}\sigma_{3}\otimes {\bf 1} + 
g_{2} \mbox{cos}(\omega_{E}t+\phi)\sigma_{1}\otimes {\bf 1}. 
\end{equation}
This model is a full generalization of (\ref{eq:hamiltonian-(ii)}), 
however we don't consider this situation in the paper. 

\vspace{3mm}\par 
We have one question : Is the Hamiltonian (\ref{eq:hamiltonian-(i)}) 
realistic or meaningful ? \quad The answer is of course yes. 
Let us show one example. 
We consider the (effective) Hamiltonian presented by NIST group 
\cite{several-1}, \cite{several-2} which were used to construct the 
controlled NOT operation (see \cite{KF1} as an introduction). 
\begin{equation}
\label{eq:hamiltonian-(iii)}
H=\omega_{0}{\bf 1}_{2}\otimes a^{\dagger}a + 
g\left(\sigma_{+}\otimes \mbox{e}^{i\eta (a^{\dagger}+a)} + 
\sigma_{-}\otimes \mbox{e}^{-i\eta (a^{\dagger}+a)}
\right) + 
\frac{\Delta}{2}\sigma_{3}\otimes {\bf 1}. 
\end{equation}

\par 
We can show that under some unitary transformation the Hamiltonian 
(\ref{eq:hamiltonian-(iii)}) can be transformed to 
(\ref{eq:hamiltonian-(i)}) with special coupling constants, 
\cite{several-3}. This is important, so we review and modify \cite{several-3}. 

We set $2A=i\eta (a^{\dagger}+a)$ for simplicity, then 
\begin{eqnarray}
&&\sigma_{+}\otimes \mbox{e}^{i\eta (a^{\dagger}+a)} + 
\sigma_{-}\otimes \mbox{e}^{-i\eta (a^{\dagger}+a)}
\nonumber \\
=&&
\left(
  \begin{array}{cc}
    0& \mbox{e}^{2A} \\
    \mbox{e}^{-2A}& 0
  \end{array}
\right)
=
\left(
  \begin{array}{cc}
    0& \mbox{e}^{A} \\
    \mbox{e}^{-A}& 0
  \end{array}
\right)
\left(
  \begin{array}{cc}
    0& 1 \\
    1& 0
  \end{array}
\right)
\left(
  \begin{array}{cc}
    0& \mbox{e}^{A} \\
    \mbox{e}^{-A}& 0  
  \end{array}
\right)  \nonumber \\
=&&
\left(
  \begin{array}{cc}
    0& \mbox{e}^{A} \\
    \mbox{e}^{-A}& 0
  \end{array}
\right)
\left\{\frac{1}{2}
\left(
  \begin{array}{cc}
    1& 1 \\
    1& -1
  \end{array}
\right)
\left(
  \begin{array}{cc}
    1& 0 \\
    0& -1
  \end{array}
\right)
\left(
  \begin{array}{cc}
    1& 1 \\
    1& -1
  \end{array}
\right)
\right\}
\left(
  \begin{array}{cc}
    0& \mbox{e}^{A} \\
    \mbox{e}^{-A}& 0  
  \end{array}
\right)  \nonumber \\
\equiv&& U(\eta)(\sigma_{3}\otimes {\bf 1})U(\eta)^{\dagger} 
\end{eqnarray}
where 
\begin{equation}
U(\eta)
=\frac{1}{\sqrt{2}}
\left(
  \begin{array}{cc}
    0& \mbox{e}^{A}   \\
    \mbox{e}^{-A}& 0
  \end{array}
\right)
\left(
  \begin{array}{cc}
    1& 1 \\
    1& -1
  \end{array}
\right)
=\frac{1}{\sqrt{2}}
\left(
  \begin{array}{cc}
    \mbox{e}^{A}& -\mbox{e}^{A}   \\
    \mbox{e}^{-A}& \mbox{e}^{-A}
  \end{array}
\right)
\end{equation}
and $\mbox{e}^{A}=D(i\eta/2)$ where $D(\beta)$ is a displacement (coherent) 
operator defined by (\ref{eq:basic-5}). 
Then it is not difficult to show 
\begin{equation}
U(\eta)^{\dagger}HU(\eta)
=\frac{\omega_0\eta^2}{4}{\bf 1}_{2}\otimes {\bf 1}+
\omega_{0}{\bf 1}_{2}\otimes a^{\dagger}a+
\frac{\omega_{0}\eta}{2}\sigma_{1}\otimes (-ia^{\dagger}+ia)+
g\sigma_{3}\otimes {\bf 1}-\frac{\Delta}{2}\sigma_{1}\otimes {\bf 1}.
\end{equation}

\par \noindent 
To remove $i$ in the term containing $a$ 
we moreover operate the unitary one 
\begin{eqnarray}
&&\left({\bf 1}_{2}\otimes \mbox{e}^{i(\pi/2) N}\right)U(\eta)^{\dagger}H
U(\eta)\left({\bf 1}_{2}\otimes \mbox{e}^{-i(\pi/2) N}\right)  \nonumber \\
=&&\frac{\omega_0\eta^2}{4}{\bf 1}_{2}\otimes {\bf 1}+
\omega_{0}{\bf 1}_{2}\otimes a^{\dagger}a+
\frac{\omega_{0}\eta}{2}\sigma_{1}\otimes (a^{\dagger}+a)+
g\sigma_{3}\otimes {\bf 1}-\frac{\Delta}{2}\sigma_{1}\otimes {\bf 1},
\end{eqnarray}
where we have used the well--known formula 
\[
\mbox{e}^{i\theta N}a\mbox{e}^{-i\theta N}=\mbox{e}^{-i\theta}a, \quad 
\mbox{e}^{i\theta N}a^{\dagger}\mbox{e}^{-i\theta N}=
\mbox{e}^{i\theta}a^{\dagger},
\]
see \cite{KF4}. Since $U(\eta)$ can be written as 
$
U(\eta)=
(\sigma_{+}\otimes \mbox{e}^{A}+\sigma_{-}\otimes \mbox{e}^{-A})
\left(W\otimes {\bf 1}\right),
$
so if we write 
\begin{equation}
T(\eta)\equiv U(\eta)({\bf 1}_{2}\otimes \mbox{e}^{-i(\pi/2) N})
=(\sigma_{+}\otimes \mbox{e}^{A}+\sigma_{-}\otimes \mbox{e}^{-A})
(W\otimes \mbox{e}^{-i(\pi/2) N}), 
\end{equation}
then we have 
\[
T(\eta)^{\dagger}HT(\eta)
=\frac{\omega_0\eta^2}{4}{\bf 1}_{2}\otimes {\bf 1}+
\omega_{0}{\bf 1}_{2}\otimes a^{\dagger}a+
\frac{\omega_{0}\eta}{2}\sigma_{1}\otimes (a^{\dagger}+a)+
g\sigma_{3}\otimes {\bf 1}-\frac{\Delta}{2}\sigma_{1}\otimes {\bf 1}. 
\]
Here we have no interest in the constant term, so we finally obtain 
\begin{equation}
\label{eq:NIST-hamiltonian}
H=T(\eta)
\left\{
\omega_{0}{\bf 1}_{2}\otimes a^{\dagger}a+
\frac{\omega_{0}\eta}{2}\sigma_{1}\otimes (a^{\dagger}+a)+
g\sigma_{3}\otimes {\bf 1}-\frac{\Delta}{2}\sigma_{1}\otimes {\bf 1}
\right\}T(\eta)^{\dagger}.
\end{equation}
$T(\eta)$ is just the unitary transformation required. We note that 
the last term is constant, which case is a special one.

\vspace{5mm}\par 
At this stage we would like to make a further generalization of  
the Hamiltonian (\ref{eq:hamiltonian-(i)}) to make wide applications 
to Quantum Computation.@

Let $\{K_{+},K_{-},K_{3}\}$ and $\{J_{+},J_{-},J_{3}\}$ be a set of 
generators of unitary representations of Lie algebras $su(1,1)$ and 
$su(2)$. They are usually constructed by making use of two harmonic 
oscillators (two--photons) $a_{1}, a_{2}$ as 
\begin{eqnarray}
  \label{eq:schwinger-boson}
  su(1,1) &:&\quad
     K_+ = {a_1}^{\dagger}{a_2}^{\dagger},\ K_- = a_2 a_1,\ 
     K_3 = {1\over2}\left({a_1}^{\dagger}a_1 + {a_2}^{\dagger}a_2  + 1\right),
  \nonumber \\
  su(2) &:&\quad
     J_+ = {a_1}^{\dagger}a_2,\ J_- = {a_2}^{\dagger}a_1,\ 
     J_3 = {1\over2}\left({a_1}^{\dagger}a_1 - {a_2}^{\dagger}a_2\right).
  \nonumber
\end{eqnarray}
Then we can make the similar arguments done for the Heisenberg 
algebra $\{a^{\dagger},a,N\}$, namely (\ref{eq:basic-1}) $\sim$ 
(\ref{eq:basic-5}), see for example \cite{KF4}. 

\par 
We have considered the following three Hamiltonians in \cite{KF2} : 
\begin{eqnarray}
\mbox{(N)}\qquad H_{N}&=&\omega {\bf 1}_{2}\otimes a^{\dagger}a + 
\frac{\Delta}{2}\sigma_{3}\otimes {\bf 1} +
g\sigma_{1}\otimes (a^{\dagger}+a), \\
\mbox{(K)}\qquad H_{K}&=&\omega {\bf 1}_{2}\otimes K_{3} + 
\frac{\Delta}{2}\sigma_{3}\otimes {\bf 1}_{K} +
g\sigma_{1}\otimes (K_{+}+K_{-}), \\
\mbox{(J)}\qquad H_{J}&=&\omega {\bf 1}_{2}\otimes J_{3} + 
\frac{\Delta}{2}\sigma_{3}\otimes {\bf 1}_{J} +
g\sigma_{1}\otimes (J_{+}+J_{-}).
\end{eqnarray}

\par \noindent 
To deal with these three cases at the same time we set 
\begin{equation}
\{L_{+},L_{-},L_{3}\}=
\left\{
\begin{array}{ll}
\mbox{(N)}\qquad \{a^{\dagger},a,N\}, \\
\mbox{(K)}\qquad \{K_{+},K_{-},K_{3}\}, \\
\mbox{(J)}\qquad \ \{J_{+},J_{-},J_{3}\} 
\end{array}
\right.
\end{equation}
and 
\begin{equation}
H_{L}=\omega {\bf 1}_{2}\otimes L_{3} + 
\frac{\Delta}{2}\sigma_{3}\otimes {\bf 1}_{L} +
g\sigma_{1}\otimes (L_{+}+L_{-}).
\end{equation}

\par \noindent 
Therefore the Hamiltonian that we are looking for is 
\vspace{3mm} \par \noindent
{\bf Unified Hamiltonian}
\begin{equation}
\label{eq:general-hamiltonian}
{\tilde H}_{L}
=\omega {\bf 1}_{2}\otimes L_{3} + 
g_{1}\sigma_{1}\otimes (L_{+}+L_{-}) + 
\frac{\Delta}{2}\sigma_{3}\otimes {\bf 1}_{L} + 
g_{2} \mbox{cos}(\omega_{E}t+\phi)\sigma_{1}\otimes {\bf 1}_{L}. 
\end{equation}

By the way, from the lesson in (\ref{eq:NIST-hamiltonian}) we can also 
consider a weak version of (\ref{eq:general-hamiltonian}) with the last term 
being constant 
\[
\label{eq:general-hamiltonian-weak}
{\tilde H}_{L}
=\omega {\bf 1}_{2}\otimes L_{3} + 
g_{1}\sigma_{1}\otimes (L_{+}+L_{-}) + 
\frac{\Delta}{2}\sigma_{3}\otimes {\bf 1}_{L} + 
g_{2}\sigma_{1}\otimes {\bf 1}_{L}, 
\]
where we have reset $g_{2}\equiv g_{2}\mbox{cos}(\phi)$ for simplicity. 
This is not a bad model as seen from (\ref{eq:NIST-hamiltonian}). This 
restricted model has been treated in \cite{KF7}. 

\par \noindent 
Now we would like to solve the Hamiltonian (\ref{eq:general-hamiltonian}), 
especially in the strong coupling regime ($g_{1}\gg \Delta$). 

Let us transform (\ref{eq:general-hamiltonian}) into 
\begin{eqnarray}
{\tilde H}_{L}
&=&{\bf 1}_{2}\otimes \omega L_{3} + 
\sigma_{1}\otimes 
\left\{g_{1}(L_{+}+L_{-}) + g_{2} \mbox{cos}(\omega_{E}t+\phi){\bf 1}_{L}
\right\} + 
\frac{\Delta}{2}\sigma_{3}\otimes {\bf 1}_{L}     \nonumber \\
&\equiv& {\tilde H}_{0}+\frac{\Delta}{2}\sigma_{3}\otimes {\bf 1}_{L}. 
\end{eqnarray}
The method to solve is almost identical to \cite{KF2}, so we give only an 
outline. By making use of the Walsh--Hadamard matrix 
(\ref{eq:2-Walsh-Hadamard}) 
\begin{eqnarray}
{\tilde H}_{0}
&=&(W\otimes {\bf 1}_{L})
\left[
{\bf 1}_{2}\otimes \omega L_{3} + 
\sigma_{3}\otimes 
\left\{g_{1}(L_{+}+L_{-})+g_{2} \mbox{cos}(\omega_{E}t+\phi){\bf 1}_{L}\right\}
\right]
(W^{-1}\otimes {\bf 1}_{L})   \nonumber \\
&=&\sum_{\lambda=\pm 1}
\left(
\ket{\lambda}\otimes \mbox{e}^{-\frac{\lambda x}{2}(L_{+}-L_{-})}
\right)
\left\{
\Omega L_{3}+\lambda g_{2} \mbox{cos}(\omega_{E}t+\phi){\bf 1}_{L}
\right\}
\left(
\bra{\lambda}\otimes \mbox{e}^{\frac{\lambda x}{2}(L_{+}-L_{-})}
\right) \nonumber 
\end{eqnarray}
where $\ket{\lambda}$ is the eigenvectors of $\sigma_{1}$ defined in 
(\ref{eq:eigenvectors of sigma}) and $\Omega,\ x$ are given as 
\begin{equation}
\label{eq:omega-x}
(\Omega,\ x)=
\left\{
\begin{array}{ll}
(N)\quad \omega,\quad \quad \quad \qquad \qquad \ x=2g_{1}/\omega, \\
(K)\quad \omega \sqrt{1-(2g_{1}/\omega)^{2}},\quad 
x=\mbox{tanh}^{-1}(2g_{1}/\omega), \\
(J)\quad \ \omega \sqrt{1+(2g_{1}/\omega)^{2}},\quad 
x=\mbox{tan}^{-1}(2g_{1}/\omega). 
\end{array}
\right.
\end{equation}
That is, we could diagonalize the Hamiltonian ${\tilde H}_{0}$. 
Its eigenvalues $\{E_{n}(t)\}$ and eigenvectors $\{\ket{\{\lambda, n\}}\}$ 
are given respectively 
\begin{equation}
\label{eq:Eigenvalues-Eigenvectors}
(E_{n}(t),\ \ket{\{\lambda, n\}})=
\left\{
\begin{array}{ll}
(N)\quad \Omega (-\frac{g^2}{\omega^2}+n)+
\lambda g_{2} \mbox{cos}(\omega_{E}t+\phi), \quad \ket{\lambda}\otimes 
\mbox{e}^{-\frac{\lambda x}{2}(a^{\dagger}-a)}\ket{n}, \\
(K)\quad \Omega (K+n)+\lambda g_{2} \mbox{cos}(\omega_{E}t+\phi),\quad \quad 
\ket{\lambda}\otimes 
\mbox{e}^{-\frac{\lambda x}{2}(K_{+}-K_{-})}\ket{K,n}, \\
(J)\quad \ \Omega (-J+n)+\lambda g_{2} \mbox{cos}(\omega_{E}t+\phi),\quad \ 
\ket{\lambda}\otimes 
\mbox{e}^{-\frac{\lambda x}{2}(J_{+}-J_{-})}\ket{J,n} \\
\end{array}
\right.
\end{equation}
for $\lambda=\pm 1$ and $n \in \futon \cup \{0\}$, where 
$E_{n}(t)\equiv E_{n}+\lambda g_{2} \mbox{cos}(\omega_{E}t+\phi)$. 
Then ${\tilde H}_{0}$ above can be written as 
\[
{\tilde H}_{0}
=\sum_{\lambda}\sum_{n}E_{n}(t)\ket{\{\lambda, n\}}\bra{\{\lambda, n\}}.
\]

\par \vspace{3mm} 
Next we would like to solve the following Schr{\"o}dinger equation : 
\begin{equation}
\label{eq:full-equation}
i\frac{d}{dt}\Psi={\tilde H}\Psi=\left({\tilde H}_{0}+
\frac{\Delta}{2}\sigma_{3}\otimes {\bf 1}_{L}\right)\Psi. 
\end{equation}
To solve this equation we appeal to the method of constant variation. 
First let us solve 
$
i\frac{d}{dt}\Psi={\tilde H}_{0}\Psi, 
$
which general solution is given by 
$
\label{eq:partial-solution}
  \Psi(t)=U_{0}(t)\Psi_{0}
$, 
where $\Psi_{0}$ is a constant state and 
\begin{equation}
\label{eq:Basic-Unitary}
U_{0}(t)=\sum_{\lambda}\sum_{n}
\mbox{e}^{-i\{tE_{n}+\lambda(g_{2}/\omega_{E})sin(\omega_{E}t+\phi)\}}
\ket{\{\lambda, n\}}\bra{\{\lambda, n\}}.
\end{equation}
The method of constant variation goes as follows. Changing like 
$
\Psi_{0} \longrightarrow \Psi_{0}(t),
$ 
we have 
\begin{equation}
\label{eq:sub-equation}
i\frac{d}{dt}\Psi_{0}
=\frac{\Delta}{2}{U_0}^{\dagger}(\sigma_{3}\otimes {\bf 1}_{L}){U_0}\Psi_{0} 
\equiv \frac{\Delta}{2}{\tilde H}_{F}\Psi_{0}
\end{equation}
after some algebra. We must solve this equation. ${\tilde H}_{F}$ is 
\begin{eqnarray}
{\tilde H}_{F}&=&
\sum_{\lambda,\mu}\sum_{m, n}
\mbox{e}^{it(E_m-E_n)+i(\lambda-\mu)(g_{2}/\omega_{E})sin(\omega_{E}t+\phi)} 
\bra{\{\lambda,m\}}(\sigma_{3}\otimes {\bf 1}_{L})\ket{\{\mu,n\}} \ 
\ket{\{\lambda, m\}}\bra{\{\mu, n\}} \nonumber \\
&=&
\sum_{\lambda}\sum_{m, n}
\mbox{e}^{ i\{t\Omega(m-n)+2\lambda(g_{2}/\omega_{E})sin(\omega_{E}t+\phi)\} } 
\braa{m}\mbox{e}^{{\lambda x}(L_{+}-L_{-})}\kett{n} \ 
\ket{\{\lambda, m\}}\bra{\{-\lambda, n\}}, 
\end{eqnarray}
where we have used $\bra{\lambda}\sigma_{3}=\bra{-\lambda}$ 
and $\kett{n}$ is respectively 
\[
\kett{n}=
\left\{
\begin{array}{ll}
(N)\qquad \ket{n}, \\
(K)\qquad \ket{K,n}, \\
(J)\qquad\ \ket{J,n}. 
\end{array}
\right.
\]
In the following we set for simplicity 
\begin{equation}
\label{eq:time-depend-F}
\Theta(t)\equiv g_{2}\frac{\mbox{sin}(\omega_{E}t+\phi)}{\omega_{E}}.
\end{equation}
Here we divide ${\tilde H}_{F}$ into two parts
$
{\tilde H}_{F}={{\tilde H}_{F}}^{'}+{{\tilde H}_{F}}^{''}
$
where 
\begin{eqnarray}
\label{eq:Second-Hamiltonian-1}
{{\tilde H}_{F}}^{'}&=&\sum_{\lambda}\sum_{n}\mbox{e}^{2i\lambda \Theta(t)}
\braa{n}\mbox{e}^{{\lambda x}(L_{+}-L_{-})}\kett{n} \ 
\ket{\{\lambda, n\}}\bra{\{-\lambda, n\}}, \\
\label{eq:Second-Hamiltonian-2}
{{\tilde H}_{F}}^{''}&=&\sum_{\lambda}
\sum_{\stackrel{\scriptstyle m,n}{m\ne n}}
\mbox{e}^{i\{t\Omega (m-n)+2\lambda \Theta(t) \}} 
\braa{m}\mbox{e}^{{\lambda x}(L_{+}-L_{-})}\kett{n} \ 
\ket{\{\lambda, m\}}\bra{\{-\lambda, n\}}.
\end{eqnarray}
Noting 
$
\braa{n}\mbox{e}^{x(L_{+}-L_{-})}\kett{n}=
\braa{n}\mbox{e}^{-x(L_{+}-L_{-})}\kett{n}
$
by the results in section 3 of \cite{KF2}, ${{\tilde H}_{F}}^{'}$ can be 
written as
\begin{equation}
\label{eq:diagonal-hamiltonian}
{{\tilde H}_{F}}^{'}=\sum_{n}\braa{n}\mbox{e}^{x(L_{+}-L_{-})}\kett{n} 
\sum_{\lambda}\mbox{e}^{2i\lambda\Theta(t)}
\ket{\{\lambda, n\}}\bra{\{-\lambda, n\}}.
\end{equation}

Here we want to solve the equation 
$i(d/dt)\Psi_{0}={{\tilde H}_{F}}^{'}\Psi_{0}$ completely, 
however it is very hard (see for example \cite{MFr1}, \cite{MFr4}, 
\cite{BaWr}, \cite{SGD}). Therefore let us appeal to a perturbation theory. 
For simplicity we set $\phi=0$ in (\ref{eq:time-depend-F}), then we have 
the well--known formula 
\begin{equation}
\mbox{e}^{2i\lambda\Theta(t)}=\sum_{\alpha\in \seisu}
J_{\alpha}(2\lambda g_{2}/\omega_{E})\mbox{e}^{i \alpha\omega_{E} t}
=J_{0}(2g_{2}/\omega_{E})+\sum_{\alpha\ne 0}
J_{\alpha}(2\lambda g_{2}/\omega_{E})\mbox{e}^{i \alpha\omega_{E} t},
\end{equation}
where $J_{\alpha}(x)$ are the Bessel functions. For a further simplicity 
we set $2g_{2}/\omega_{E}=\Gamma$.

We decompose (\ref{eq:diagonal-hamiltonian}) as 
\[
{{\tilde H}_{F}}^{'}={{\tilde H}_{0F}}^{'}+{{\tilde H}_{1F}}^{'}\ ; 
\]
where 
\begin{equation}
\label{eq:diagonal-hamiltonian-0}
{{\tilde H}_{0F}}^{'}=
\sum_{n}\braa{n}\mbox{e}^{x(L_{+}-L_{-})}\kett{n}J_{0}(\Gamma)
\sum_{\lambda}
\ket{\{\lambda, n\}}\bra{\{-\lambda, n\}}
\end{equation}
and
\begin{equation}
\label{eq:diagonal-hamiltonian-1}
{{\tilde H}_{1F}}^{'}=
\sum_{n}\braa{n}\mbox{e}^{x(L_{+}-L_{-})}\kett{n}
\sum_{\lambda}\sum_{\alpha\ne 0}
J_{\alpha}(\lambda\Gamma)\mbox{e}^{i \alpha\omega_{E} t}
\ket{\{\lambda, n\}}\bra{\{-\lambda, n\}}.
\end{equation}

\par \noindent
Next let us transform (\ref{eq:Second-Hamiltonian-2}). 
\begin{eqnarray}
\label{eq:non-diagonal-hamiltonian-01}
{{\tilde H}_{F}}^{''}
&=&
\sum_{\stackrel{\scriptstyle m,n}{m\ne n}}
\mbox{e}^{it\Omega(m-n)} 
\sum_{\lambda}\mbox{e}^{2i\lambda\Theta(t)}
\braa{m}\mbox{e}^{{\lambda x}(L_{+}-L_{-})}\kett{n} \ 
\ket{\{\lambda, m\}}\bra{\{-\lambda, n\}} \nonumber \\
&=&
\sum_{\stackrel{\scriptstyle m,n}{m\ne n}}
\mbox{e}^{it\Omega(m-n)} 
\sum_{\lambda}\sum_{\alpha}
J_{\alpha}(\lambda\Gamma)\mbox{e}^{i\alpha\omega_{E} t}
\braa{m}\mbox{e}^{{\lambda x}(L_{+}-L_{-})}\kett{n}\ 
\ket{\{\lambda, m\}}\bra{\{-\lambda, n\}}.
\end{eqnarray}

Now we define a new basis called Schr{\"o}dinger cat states 
\[
|\{\sigma,{\psi}_{n}\}\rangle = \frac{1}{\sqrt{2}}
(\ket{\{1, n\}} + \sigma\ket{\{-1, n\}}), 
\qquad \sigma=\pm 1. 
\]
From these we have easily 
\[
\ket{\{\lambda, n\}}=\frac{1}{\sqrt{2}}
(|\{1,{\psi}_{n}\}\rangle + \lambda|\{-1,{\psi}_{n}\}\rangle ), 
\qquad \lambda=\pm 1. 
\]

Now we are in a position to rewrite (\ref{eq:diagonal-hamiltonian-0}), 
(\ref{eq:diagonal-hamiltonian-1}), (\ref{eq:non-diagonal-hamiltonian-01}) 
in terms of $|\{\lambda,{\psi}_{n}\}\rangle$ : 
\begin{equation}
\label{eq:diagonal-hamiltonian-0-change}
{{\tilde H}_{0F}}^{'}=
\sum_{n}\braa{n}\mbox{e}^{x(L_{+}-L_{-})}\kett{n}J_{0}(\Gamma)
\sum_{\lambda}\lambda
\ket{\{\lambda,{\psi}_{n}\}}\bra{\{\lambda,{\psi}_{n}\}}
\end{equation}
and
\begin{eqnarray}
\label{eq:diagonal-hamiltonian-1-change}
{{\tilde H}_{1F}}^{'}&=&
\sum_{n}\braa{n}\mbox{e}^{x(L_{+}-L_{-})}\kett{n}
\sum_{\alpha\ne 0}\mbox{e}^{i \alpha\omega_{E} t}
\sum_{\sigma}J_{\alpha}(\sigma\Gamma)\times  \nonumber \\
&&\frac{1}{2}
\left\{
\sum_{\lambda}\lambda
\ket{\{\lambda,{\psi}_{n}\}}\bra{\{\lambda,{\psi}_{n}\}}-
\sigma\sum_{\lambda}\lambda
\ket{\{\lambda,{\psi}_{n}\}}\bra{\{-\lambda,{\psi}_{n}\}}
\right\},   
\end{eqnarray}
and moreover 
\begin{eqnarray}
\label{eq:non-diagonal-hamiltonian-01-change}
{{\tilde H}_{F}}^{''}&=&
\sum_{\stackrel{\scriptstyle m,n}{m\ne n}}
\mbox{e}^{it\Omega(m-n)} 
\sum_{\alpha}\mbox{e}^{i\alpha\omega_{E} t}
\sum_{\sigma}
J_{\alpha}(\sigma\Gamma)\braa{m}\mbox{e}^{{\sigma x}(L_{+}-L_{-})}\kett{n}
\times  \nonumber \\
&&\frac{1}{2}
\left\{
\sum_{\lambda}\lambda \ket{\{\lambda,{\psi}_{m}\}}\bra{\{\lambda,{\psi}_{n}\}}
-\sigma \sum_{\lambda}\lambda 
\ket{\{\lambda,{\psi}_{m}\}}\bra{\{-\lambda,{\psi}_{n}\}}
\right\}.
\end{eqnarray}

\par \noindent
For simplicity in the following we set 
\begin{equation}
E_{\Delta,n,\lambda}=\frac{\Delta}{2}\lambda
\braa{n}\mbox{e}^{x(L_{+}-L_{-})}\kett{n}J_{0}(\Gamma), 
\quad \Gamma=\frac{2g_{2}}{\omega_{E}}
\end{equation}
then 
\begin{equation}
E_{\Delta,n,\lambda}=
\left\{
\begin{array}{ll}
(N)\quad \frac{\Delta}{2}\lambda
\mbox{e}^{-\frac{\kappa^2}{2}}L_{n}\left(\kappa^{2}\right)J_{0}(\Gamma)
\quad  \mbox{where}\quad \kappa=x  \\
(K)\quad \frac{\Delta}{2}\lambda
\frac{n!}{(2K)_{n}}(1+\kappa^2)^{-K-n}
F_{n}(\kappa^2:2K)J_{0}(\Gamma)
\quad \mbox{where}\quad \kappa=\mbox{sinh}(x)  \\
(J)\quad \frac{\Delta}{2}\lambda
\frac{n!}{{}_{2J}P_n}(1-\kappa^2)^{J-n}
F_{n}(\kappa^2:2J)J_{0}(\Gamma)
\quad \mbox{where}\quad \kappa=\mbox{sin}(x)  
\end{array}
\right.
\end{equation}
from the results in sectin 3.1 of \cite{KF2}.

Then we rewrite (\ref{eq:sub-equation}) as 
\begin{equation}
\label{eq:sub-equation-modify}
i\frac{d}{dt}\Psi_{0}=\frac{\Delta}{2}
\left\{
{{\tilde H}_{0F}}^{'}+
\left({{\tilde H}_{1F}}^{'}+{{\tilde H}_{F}}^{''} \right)
\right\}
\Psi_{0}. 
\end{equation}
and appeal to a perturbation method. 
Namely, we treat ${{\tilde H}_{0F}}^{'}$ a unperturbed Hamiltonian 
and the remainig a perturbed one. 

\par \noindent 
It is easy to solve the equation 
$i(d/dt)\Psi_{0}=(\Delta/2){{\tilde H}_{0F}}^{'}\Psi_{0}$, 
which solution is given by 
\[
\Psi_{0}=
\sum_{n}\sum_{\lambda}
\mbox{e}^{-itE_{\Delta,n,\lambda}}c_{n,\lambda}\ket{\{\lambda,{\psi}_{n}\}},
\]
where $\{c_{n,\lambda} \}$ are constant, 
so we can set an ansatz to solve (\ref{eq:sub-equation-modify}) as 
\begin{equation}
\label{eq:ansatz}
\Psi_{0}=
\sum_{n}\sum_{\lambda}
\mbox{e}^{-itE_{\Delta,n,\lambda}}c_{n,\lambda}(t)
\ket{\{\lambda,{\psi}_{n}\}}
\end{equation}
and determine the coefficients $\{c_{n,\lambda}(t) \}$ from 
(\ref{eq:sub-equation-modify}). 
However the (infinite) equations are almost impossible to solve. On the 
other hand we are interested in Quantum Computation, so let us restrict the 
ansatz (\ref{eq:ansatz}) which is enough for our purpose. 
\begin{flushleft}
{\sl {\Large One Qubit Case}}
\end{flushleft}
The ansatz is simple 
\begin{equation}
\label{eq:ansatz-one}
\Psi_{0}=\sum_{\lambda\in \{1,-1\}}
\mbox{e}^{-itE_{\Delta,n,\lambda}}c_{n,\lambda}(t)\ket{\{\lambda,{\psi}_{n}\}},
\end{equation}
where $n$ is fixed. Substituting this into (\ref{eq:sub-equation-modify}) and 
after some algebras we obtain 
\begin{eqnarray}
&&i\mbox{e}^{-itE_{\Delta,n,\lambda}}\frac{d}{dt}c_{n,\lambda}(t)=
\nonumber \\
&&\frac{\Delta}{2}\lambda \braa{n}\mbox{e}^{x(L_{+}-L_{-})}\kett{n} 
\sum_{\alpha\ne 0}\mbox{e}^{i\alpha\omega_{E} t}
\sum_{\sigma}\frac{J_{\alpha}(\sigma\Gamma)}{2}
\left(
\mbox{e}^{-itE_{\Delta,n,\lambda}}c_{n,\lambda}(t)-\sigma
\mbox{e}^{-itE_{\Delta,n,-\lambda}}c_{n,-\lambda}(t)
\right), \nonumber 
\end{eqnarray}
so we have 
\begin{eqnarray}
&&\frac{d}{dt}c_{n,\lambda}(t)=      \nonumber \\
&&{-i}\frac{\Delta}{2}\lambda \braa{n}\mbox{e}^{x(L_{+}-L_{-})}\kett{n} 
\sum_{\alpha\ne 0}
\sum_{\sigma}\frac{J_{\alpha}(\sigma\Gamma)}{2}
\left(
\mbox{e}^{i\alpha\omega_{E} t}c_{n,\lambda}(t)-\sigma
\mbox{e}^{it(\alpha\omega_{E}-2E_{\Delta,n,-\lambda})}c_{n,-\lambda}(t)
\right). \nonumber 
\end{eqnarray}

Now we set a {\bf resonance condition} : for some $\alpha \ne 0 \in \seisu$ 
\begin{eqnarray}
\label{eq:resonance}
\alpha\omega_{E}-2E_{\Delta,n,-1}=0
\ \Longleftrightarrow \ 
(-\alpha)\omega_{E}-2E_{\Delta,n,1}=0,
\end{eqnarray}
because $E_{\Delta,n,-\lambda}=-E_{\Delta,n,\lambda}$. 
Then the remaining terms might be neglected (a kind of {\bf rotating wave 
approximation}), so we have 
\begin{eqnarray}
\frac{d}{dt}c_{n,1}(t)&=& 
{i}\frac{\Delta}{2} \braa{n}\mbox{e}^{x(L_{+}-L_{-})}\kett{n} 
\sum_{\sigma}\sigma \frac{J_{\alpha}(\sigma\Gamma)}{2}c_{n,-1}(t),
\nonumber \\
\frac{d}{dt}c_{n,-1}(t)&=& 
{-i}\frac{\Delta}{2} \braa{n}\mbox{e}^{x(L_{+}-L_{-})}\kett{n} 
\sum_{\sigma}\sigma \frac{J_{-\alpha}(\sigma\Gamma)}{2}c_{n,1}(t).
\nonumber 
\end{eqnarray}
By the way, from the fact 
$J_{-\alpha}(x)=(-1)^{\alpha}J_{\alpha}(x)=J_{\alpha}(-x)$
we have 
\[
\sum_{\sigma}\sigma J_{-\alpha}(\sigma\Gamma)
=
\sum_{\sigma}\sigma J_{\alpha}(-\sigma\Gamma)
=
\sum_{\sigma}(-\sigma) J_{\alpha}(\sigma\Gamma)
=
-\sum_{\sigma}\sigma J_{\alpha}(\sigma\Gamma).
\]
Therefore the equations above become 
\begin{equation}
\label{eq:reduced-equation}
\frac{d}{dt}
\left(
  \begin{array}{c}
    c_{n,1} \\
    c_{n,-1}
  \end{array}
\right)
=
\left(
  \begin{array}{cc}
     0& i\frac{{\cal R}}{2}   \\
     i\frac{{\cal R}}{2} & 0
  \end{array}
\right)
\left(
  \begin{array}{c}
    c_{n,1} \\
    c_{n,-1}
  \end{array}
\right)
\end{equation}
with 
\begin{equation}
{\cal R}=
{\Delta} \braa{n}\mbox{e}^{x(L_{+}-L_{-})}\kett{n} 
\sum_{\sigma}\sigma \frac{J_{\alpha}(\sigma\Gamma)}{2}
=
{\Delta} \braa{n}\mbox{e}^{x(L_{+}-L_{-})}\kett{n}J_{\alpha}(\Gamma)
\frac{1-(-1)^{\alpha}}{2}. 
\end{equation}
We note that this ${\cal R}$ is just the Rabi (flopping) frequency. 
The solution is given by 
\begin{equation}
\label{eq:1-bit-solution}
\left(
  \begin{array}{c}
    c_{n,1}(t) \\
    c_{n,-1}(t)
  \end{array}
\right)
=
\left(
  \begin{array}{cc}
    \mbox{cos}(\frac{{\cal R}}{2}t) & i\mbox{sin}(\frac{{\cal R}}{2}t) \\
    i\mbox{sin}(\frac{{\cal R}}{2}t) & \mbox{cos}(\frac{{\cal R}}{2}t)
  \end{array}
\right)
\left(
  \begin{array}{c}
    c_{n,1}(0) \\
    c_{n,-1}(0)
  \end{array}
\right). 
\end{equation}

\vspace{5mm}
\begin{flushleft}
{\sl {\Large Two Qubit Case}}
\end{flushleft}
We identify the two--qubit space with two excited states, \cite{several-1}, 
\cite{several-2}, \cite{KF2}. 

For $m < n$ the ansatz is 
\begin{equation}
\label{eq:ansatz-two}
\Psi_{0}=
\sum_{\lambda \in \{1,-1\}}
\mbox{e}^{-itE_{\Delta,m,\lambda}}c_{m,\lambda}(t)\ket{\{\lambda,{\psi}_{m}\}}
+
\sum_{\lambda \in \{1,-1\}}
\mbox{e}^{-itE_{\Delta,n,\lambda}}c_{n,\lambda}(t)\ket{\{\lambda,{\psi}_{n}\}}.
\end{equation}
Substituting this into (\ref{eq:sub-equation-modify}) and after long algebras 
we obtain 
\begin{eqnarray}
&&i\mbox{e}^{-itE_{\Delta,m,\lambda}}\frac{d}{dt}c_{m,\lambda}(t)=
\nonumber \\
&&\frac{\Delta}{2}\lambda \braa{m}\mbox{e}^{x(L_{+}-L_{-})}\kett{m} 
\sum_{\alpha\ne 0}\mbox{e}^{i\alpha\omega_{E} t}
\sum_{\sigma}\frac{J_{\alpha}(\sigma\Gamma)}{2}
\left(
\mbox{e}^{-itE_{\Delta,m,\lambda}}c_{m,\lambda}(t)-\sigma
\mbox{e}^{-itE_{\Delta,m,-\lambda}}c_{m,-\lambda}(t)
\right) + \nonumber \\
&&\frac{\Delta}{2}\lambda\mbox{e}^{it\Omega(m-n)}
\sum_{\alpha}\mbox{e}^{i\alpha\omega_{E} t}
\sum_{\sigma}\frac{J_{\alpha}(\sigma\Gamma)}{2}
\braa{m}\mbox{e}^{\sigma x(L_{+}-L_{-})}\kett{n} 
\left(
\mbox{e}^{-itE_{\Delta,n,\lambda}}c_{n,\lambda}(t)-
\sigma\mbox{e}^{-itE_{\Delta,n,-\lambda}}c_{n,-\lambda}(t)
\right), \nonumber 
\end{eqnarray}
\begin{eqnarray}
&&i\mbox{e}^{-itE_{\Delta,n,\lambda}}\frac{d}{dt}c_{n,\lambda}(t)=
\nonumber \\
&&\frac{\Delta}{2}\lambda \braa{n}\mbox{e}^{x(L_{+}-L_{-})}\kett{n} 
\sum_{\alpha\ne 0}\mbox{e}^{i\alpha\omega_{E} t}
\sum_{\sigma}\frac{J_{\alpha}(\sigma\Gamma)}{2}
\left(
\mbox{e}^{-itE_{\Delta,n,\lambda}}c_{n,\lambda}(t)-\sigma
\mbox{e}^{-itE_{\Delta,n,-\lambda}}c_{n,-\lambda}(t)
\right) + \nonumber \\
&&\frac{\Delta}{2}\lambda\mbox{e}^{it\Omega(n-m)}
\sum_{\alpha}\mbox{e}^{i\alpha\omega_{E} t}
\sum_{\sigma}\frac{J_{\alpha}(\sigma\Gamma)}{2}
\braa{n}\mbox{e}^{\sigma x(L_{+}-L_{-})}\kett{m} 
\left(
\mbox{e}^{-itE_{\Delta,m,\lambda}}c_{m,\lambda}(t)-
\sigma\mbox{e}^{-itE_{\Delta,m,-\lambda}}c_{m,-\lambda}(t)
\right). \nonumber 
\end{eqnarray}
Then we have 
\begin{eqnarray}
&&i\frac{d}{dt}c_{m,\lambda}(t)=            \nonumber \\
&&\frac{\Delta}{2}\lambda \braa{m}\mbox{e}^{x(L_{+}-L_{-})}\kett{m} 
\sum_{\alpha\ne 0}
\sum_{\sigma}\frac{J_{\alpha}(\sigma\Gamma)}{2}
\left(\mbox{e}^{i\alpha\omega_{E} t}c_{m,\lambda}(t)-
\sigma\mbox{e}^{it(\alpha\omega_{E}-2E_{\Delta,m,-\lambda})}c_{m,-\lambda}(t)
\right) + \nonumber \\
&&\frac{\Delta}{2}\lambda
\sum_{\alpha}
\sum_{\sigma}\frac{J_{\alpha}(\sigma\Gamma)}{2}
\braa{m}\mbox{e}^{\sigma x(L_{+}-L_{-})}\kett{n}\times   \nonumber \\
&&\left\{
\mbox{e}^{it\left(\alpha\omega_{E}+\Omega(m-n)+E_{\Delta,m,\lambda}
-E_{\Delta,n,\lambda}\right)}c_{n,\lambda}(t)-
\sigma\mbox{e}^{it\left(\alpha\omega_{E}+\Omega(m-n)+E_{\Delta,m,\lambda}
-E_{\Delta,n,-\lambda}\right)}c_{n,-\lambda}(t)
\right\}, \nonumber 
\end{eqnarray}
\begin{eqnarray}
&&i\frac{d}{dt}c_{n,\lambda}(t)=
\nonumber \\
&&\frac{\Delta}{2}\lambda \braa{n}\mbox{e}^{x(L_{+}-L_{-})}\kett{n} 
\sum_{\alpha\ne 0}
\sum_{\sigma}\frac{J_{\alpha}(\sigma\Gamma)}{2}
\left(\mbox{e}^{i\alpha\omega_{E} t}c_{n,\lambda}(t)-
\sigma\mbox{e}^{it(\alpha\omega_{E}-2E_{\Delta,n,-\lambda})}c_{n,-\lambda}(t)
\right) + \nonumber \\
&&\frac{\Delta}{2}\lambda
\sum_{\alpha}
\sum_{\sigma}\frac{J_{\alpha}(\sigma\Gamma)}{2}
\braa{n}\mbox{e}^{\sigma x(L_{+}-L_{-})}\kett{m}\times   \nonumber \\
&&\left\{
\mbox{e}^{it\left(\alpha\omega_{E}+\Omega(n-m)+E_{\Delta,n,\lambda}
-E_{\Delta,m,\lambda}\right)}c_{m,\lambda}(t)-
\sigma\mbox{e}^{it\left(\alpha\omega_{E}+\Omega(n-m)+E_{\Delta,n,\lambda}
-E_{\Delta,m,-\lambda}\right)}c_{m,-\lambda}(t)
\right\}. \nonumber 
\end{eqnarray}

\par \noindent
Now we can set four (possible) resonance conditions : 

\par \noindent
(1)\quad a {\bf resonance condition} : for some $\alpha \in \seisu$ 
\begin{eqnarray}
\label{eq:resonance-1}
&&\alpha\omega_{E}+\Omega(m-n)+E_{\Delta,m,1}-E_{\Delta,n,1}=0
\nonumber \\
\Longleftrightarrow\ 
&&(-\alpha)\omega_{E}+\Omega(n-m)+E_{\Delta,n,1}-E_{\Delta,m,1}=0.
\end{eqnarray}
Then by using the rotating wave approximation we obtain the equations 
\begin{equation}
\label{eq:reduced-equation-1}
\frac{d}{dt}
\left(
  \begin{array}{c}
    c_{m,1}  \\
    c_{m,-1} \\
    c_{n,1}  \\
    c_{n,-1}
  \end{array}
\right)
=
\left(
  \begin{array}{cccc}
     0& &-i\frac{{\cal R}}{2}& \\
      &0& &                    \\
    -i\frac{{\bar{\cal R}}}{2}& &0 &  \\
      & & & 0
  \end{array}
\right)
\left(
  \begin{array}{c}
    c_{m,1}  \\
    c_{m,-1} \\
    c_{n,1}  \\
    c_{n,-1}
  \end{array}
\right)
\end{equation}
with 
\begin{eqnarray}
{\cal R}
&=&{\Delta}\sum_{\sigma}\frac{J_{\alpha}(\sigma\Gamma)}{2}
 \braa{m}\mbox{e}^{\sigma x(L_{+}-L_{-})}\kett{n}    \nonumber \\
&=&{\Delta}J_{\alpha}(\Gamma)
\frac{1}{2}
\left\{
\braa{m}\mbox{e}^{x(L_{+}-L_{-})}\kett{n}+
(-1)^{\alpha}\braa{m}\mbox{e}^{-x(L_{+}-L_{-})}\kett{n}
\right\}. 
\end{eqnarray}
Here we have used the following identity : neglecting $\Delta$ in ${\cal R}$ 
\begin{eqnarray}
{\bar{\cal R}}&=&
\left\{
\sum_{\sigma}\frac{J_{\alpha}(\sigma\Gamma)}{2}
\braa{m}\mbox{e}^{\sigma x(L_{+}-L_{-})}\kett{n}
\right\}^{\dagger}  
=
\sum_{\sigma}\frac{J_{\alpha}(\sigma\Gamma)}{2}
\braa{n}\mbox{e}^{-\sigma x(L_{+}-L_{-})}\kett{m}  \nonumber \\
&=&
\sum_{\sigma}\frac{J_{\alpha}(-\sigma\Gamma)}{2}
\braa{n}\mbox{e}^{\sigma x(L_{+}-L_{-})}\kett{m}  
=
\sum_{\sigma}\frac{J_{-\alpha}(\sigma\Gamma)}{2}
\braa{n}\mbox{e}^{\sigma x(L_{+}-L_{-})}\kett{m}.  \nonumber 
\end{eqnarray}
For the explicit value for matrix element 
$\braa{m}\mbox{e}^{\sigma x(L_{+}-L_{-})}\kett{n}$ ($\sigma = \pm 1$) 
see \cite{KF2}. 

\par \noindent 
The solution is 
\begin{equation}
\label{eq:solution-1}
\left(
  \begin{array}{c}
    c_{m,1}(t)  \\
    c_{m,-1}(t) \\
    c_{n,1}(t)  \\
    c_{n,-1}(t)
  \end{array}
\right)
=
\left(
  \begin{array}{cccc}
   \mbox{cos}(t\frac{|{\cal R}|}{2})& &
   -i\frac{{\cal R}}{|{\cal R}|}\mbox{sin}(t\frac{|{\cal R}|}{2})& \\
   &1& &   \\
   -i\frac{{\bar{\cal R}}}{|{\cal R}|}\mbox{sin}(t\frac{|{\cal R}|}{2})& &
    \mbox{cos}(t\frac{|{\cal R}|}{2})&  \\
      & & & 1
  \end{array}
\right)
\left(
  \begin{array}{c}
    c_{m,1}(0)  \\
    c_{m,-1}(0) \\
    c_{n,1}(0)  \\
    c_{n,-1}(0)
  \end{array}
\right). 
\end{equation}

\par \noindent 
(2)\quad a {\bf resonance condition} : for some $\alpha \in \seisu$ 
\begin{eqnarray}
\label{eq:resonance-2}
&&\alpha\omega_{E}+\Omega(m-n)+E_{\Delta,m,-1}-E_{\Delta,n,-1}=0
\nonumber \\
\Longleftrightarrow\ 
&&(-\alpha)\omega_{E}+\Omega(n-m)+E_{\Delta,n,-1}-E_{\Delta,m,-1}=0.
\end{eqnarray}
Then by using the rotating wave approximation we obtain the equations 
\begin{equation}
\label{eq:reduced-equation-1}
\frac{d}{dt}
\left(
  \begin{array}{c}
    c_{m,1}  \\
    c_{m,-1} \\
    c_{n,1}  \\
    c_{n,-1}
  \end{array}
\right)
=
\left(
  \begin{array}{cccc}
     0& & & \\
      &0& & i\frac{{\cal R}}{2}      \\
      & &0 &  \\
      & i\frac{{\bar{\cal R}}}{2}& & 0
  \end{array}
\right)
\left(
  \begin{array}{c}
    c_{m,1}  \\
    c_{m,-1} \\
    c_{n,1}  \\
    c_{n,-1}
  \end{array}
\right)
\end{equation}
with 
\begin{equation}
{\cal R}={\Delta}
\sum_{\sigma}\frac{J_{\alpha}(\sigma\Gamma)}{2}
\braa{m}\mbox{e}^{\sigma x(L_{+}-L_{-})}\kett{n}.
\end{equation}
The solution is 
\begin{equation}
\label{eq:solution-1}
\left(
  \begin{array}{c}
    c_{m,1}(t)  \\
    c_{m,-1}(t) \\
    c_{n,1}(t)  \\
    c_{n,-1}(t)
  \end{array}
\right)
=
\left(
  \begin{array}{cccc}
   1& & & \\
   &\mbox{cos}(t\frac{|{\cal R}|}{2})& 
   &i\frac{{\cal R}}{|{\cal R}|}\mbox{sin}(t\frac{|{\cal R}|}{2})  \\
   & & 1&  \\
   &i\frac{{\bar{\cal R}}}{|{\cal R}|}\mbox{sin}(t\frac{|{\cal R}|}{2}) & 
   & \mbox{cos}(t\frac{|{\cal R}|}{2})
  \end{array}
\right)
\left(
  \begin{array}{c}
    c_{m,1}(0)  \\
    c_{m,-1}(0) \\
    c_{n,1}(0)  \\
    c_{n,-1}(0)
  \end{array}
\right). 
\end{equation}

\par \noindent
(3)\quad a {\bf resonance condition} : for some $\alpha \in \seisu$ 
\begin{eqnarray}
\label{eq:resonance-3}
&&\alpha\omega_{E}+\Omega(m-n)+E_{\Delta,m,1}-E_{\Delta,n,-1}=0
\nonumber \\
\Longleftrightarrow\ 
&&(-\alpha)\omega_{E}+\Omega(n-m)+E_{\Delta,n,-1}-E_{\Delta,m,1}=0.
\end{eqnarray}
Then by using the rotating wave approximation we obtain the equations 
\begin{equation}
\label{eq:reduced-equation-1}
\frac{d}{dt}
\left(
  \begin{array}{c}
    c_{m,1}  \\
    c_{m,-1} \\
    c_{n,1}  \\
    c_{n,-1}
  \end{array}
\right)
=
\left(
  \begin{array}{cccc}
     0& & & i\frac{{\cal R}}{2}  \\
     &0 & &         \\
     &  &0 &  \\
    i\frac{{\bar{\cal R}}}{2}& & &0
  \end{array}
\right)
\left(
  \begin{array}{c}
    c_{m,1}  \\
    c_{m,-1} \\
    c_{n,1}  \\
    c_{n,-1}
  \end{array}
\right)
\end{equation}
with 
\begin{eqnarray}
{\cal R}
&=&{\Delta}\sum_{\sigma}\sigma \frac{J_{\alpha}(\sigma\Gamma)}{2}
 \braa{m}\mbox{e}^{\sigma x(L_{+}-L_{-})}\kett{n}    \nonumber \\
&=&{\Delta}J_{\alpha}(\Gamma)\frac{1}{2}
\left\{
 \braa{m}\mbox{e}^{x(L_{+}-L_{-})}\kett{n}-
 (-1)^{\alpha}\braa{m}\mbox{e}^{-x(L_{+}-L_{-})}\kett{n}
\right\}.
\end{eqnarray}
Here we have used the following identity : neglecting $\Delta$ in ${\cal R}$ 
\begin{eqnarray}
{\bar{\cal R}}&=&
\left\{
\sum_{\sigma}\sigma \frac{J_{\alpha}(\sigma\Gamma)}{2}
\braa{m}\mbox{e}^{\sigma x(L_{+}-L_{-})}\kett{n}
\right\}^{\dagger}  
=
\sum_{\sigma}\sigma \frac{J_{\alpha}(\sigma\Gamma)}{2}
\braa{n}\mbox{e}^{-\sigma x(L_{+}-L_{-})}\kett{m}     \nonumber \\
&=&
\sum_{\sigma}(-\sigma)\frac{J_{\alpha}(-\sigma\Gamma)}{2}
\braa{n}\mbox{e}^{\sigma x(L_{+}-L_{-})}\kett{m} 
=
-\sum_{\sigma}\sigma \frac{J_{-\alpha}(\sigma\Gamma)}{2}
\braa{n}\mbox{e}^{\sigma x(L_{+}-L_{-})}\kett{m}, 
\end{eqnarray}
so 
\begin{equation}
\sum_{\sigma}\sigma \frac{J_{-\alpha}(\sigma\Gamma)}{2}
\braa{n}\mbox{e}^{\sigma x(L_{+}-L_{-})}\kett{m}=-{\bar{\cal R}}. 
\end{equation}
The solution is 
\begin{equation}
\label{eq:solution-1}
\left(
  \begin{array}{c}
    c_{m,1}(t)  \\
    c_{m,-1}(t) \\
    c_{n,1}(t)  \\
    c_{n,-1}(t)
  \end{array}
\right)
=
\left(
  \begin{array}{cccc}
   \mbox{cos}(t\frac{|{\cal R}|}{2})& & 
   &i\frac{{\cal R}}{|{\cal R}|}\mbox{sin}(t\frac{|{\cal R}|}{2})  \\
   &1 & &  \\
   & &1 &  \\
   i\frac{{\bar{\cal R}}}{|{\cal R}|}\mbox{sin}(t\frac{|{\cal R}|}{2})& & 
   & \mbox{cos}(t\frac{|{\cal R}|}{2})
  \end{array}
\right)
\left(
  \begin{array}{c}
    c_{m,1}(0)  \\
    c_{m,-1}(0) \\
    c_{n,1}(0)  \\
    c_{n,-1}(0)
  \end{array}
\right). 
\end{equation}

\par \noindent
(4)\quad a {\bf resonance condition} : for some $\alpha \in \seisu$ 
\begin{eqnarray}
\label{eq:resonance-3}
&&\alpha\omega_{E}+\Omega(m-n)+E_{\Delta,m,-1}-E_{\Delta,n,1}=0
\nonumber \\
\Longleftrightarrow\ 
&&(-\alpha)\omega_{E}+\Omega(n-m)+E_{\Delta,n,1}-E_{\Delta,m,-1}=0.
\end{eqnarray}
Then by using the rotating wave approximation we obtain the equations 
\begin{equation}
\label{eq:reduced-equation-1}
\frac{d}{dt}
\left(
  \begin{array}{c}
    c_{m,1}  \\
    c_{m,-1} \\
    c_{n,1}  \\
    c_{n,-1}
  \end{array}
\right)
=
\left(
  \begin{array}{cccc}
     0& & &  \\
     &0 & -i\frac{{\cal R}}{2} &        \\
     &-i\frac{{\bar{\cal R}}}{2}  &0 &  \\
     & & &0
  \end{array}
\right)
\left(
  \begin{array}{c}
    c_{m,1}  \\
    c_{m,-1} \\
    c_{n,1}  \\
    c_{n,-1}
  \end{array}
\right)
\end{equation}
with 
\begin{equation}
{\cal R}={\Delta}
\sum_{\sigma}\sigma \frac{J_{\alpha}(\sigma\Gamma)}{2}
\braa{m}\mbox{e}^{\sigma x(L_{+}-L_{-})}\kett{n}.
\end{equation}
The solution is 
\begin{equation}
\label{eq:solution-1}
\left(
  \begin{array}{c}
    c_{m,1}(t)  \\
    c_{m,-1}(t) \\
    c_{n,1}(t)  \\
    c_{n,-1}(t)
  \end{array}
\right)
=
\left(
  \begin{array}{cccc}
   1& & &  \\
   &\mbox{cos}(t\frac{|{\cal R}|}{2}) 
   &-i\frac{{\cal R}}{|{\cal R}|}\mbox{sin}(t\frac{|{\cal R}|}{2}) &  \\
   &-i\frac{{\bar{\cal R}}}{|{\cal R}|}\mbox{sin}(t\frac{|{\cal R}|}{2}) 
   & \mbox{cos}(t\frac{|{\cal R}|}{2})&  \\
   & & & 1
  \end{array}
\right)
\left(
  \begin{array}{c}
    c_{m,1}(0)  \\
    c_{m,-1}(0) \\
    c_{n,1}(0)  \\
    c_{n,-1}(0)
  \end{array}
\right). 
\end{equation}

\par \vspace{10mm}
On the ansatz (\ref{eq:ansatz-two}) 
we solved the Schr{\"o}dinger equation (\ref{eq:sub-equation-modify}) in the 
strong coupling regime (!) under the resonance conditions and rotating wave 
approximations, and obtained the unitary transformations of four types 
which are a generalization of \cite{KF2}. They will play an important role in 
not only Quantum Computation but also Quantum Optics or Condensed Matter 
Physics. 

\par \noindent
It is very interesting that each Rabi frequency ${\cal R}$ contains some 
Bessel functions $\{J_{\alpha}(\Gamma)\}$. 
See \cite{NPT} for an interesting experiment which the Bessel function 
$J_{0}$ appeared. See also \cite{MFr2}. 

By the way, we considered the case of one atom with two--level, so we would 
like to generalize our method to the case of $n$ atoms (with two--level) 
interacting both the single radiation mode and external periodic fields 
like ($n$ atoms trapped in a cavity with the photon interaction) 
\vspace{5mm} 
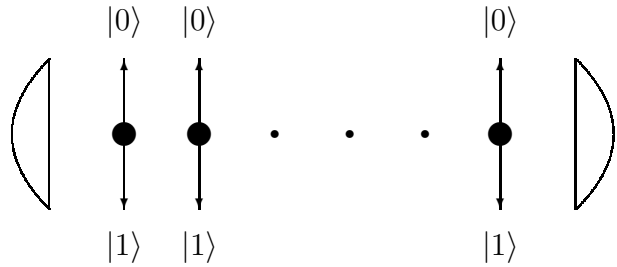
\begin{center}
\setlength{\unitlength}{1mm} 
\begin{picture}(120,40)(0,-10)
\bezier{200}(20,0)(10,10)(20,20)
\put(20,0){\line(0,1){20}}
\put(20,20){\makebox(20,10)[c]{$|0\rangle$}}
\put(30,10){\vector(0,1){10}}
\put(30,10){\vector(0,-1){10}}
\put(20,-10){\makebox(20,10)[c]{$|1\rangle$}}
\put(30,10){\circle*{3}}
\put(30,20){\makebox(20,10)[c]{$|0\rangle$}}
\put(40,10){\vector(0,1){10}}
\put(40,10){\vector(0,-1){10}}
\put(30,-10){\makebox(20,10)[c]{$|1\rangle$}}
\put(40,10){\circle*{3}}
\put(50,10){\circle*{1}}
\put(60,10){\circle*{1}}
\put(70,10){\circle*{1}}
\put(70,20){\makebox(20,10)[c]{$|0\rangle$}}
\put(80,10){\vector(0,1){10}}
\put(80,10){\vector(0,-1){10}}
\put(80,10){\circle*{3}}
\put(70,-10){\makebox(20,10)[c]{$|1\rangle$}}
\bezier{200}(90,0)(100,10)(90,20)
\put(90,0){\line(0,1){20}}
\end{picture}
\end{center}
Then the Hamiltonian may be 
\begin{equation}
\label{eq:general-n-hamiltonian}
{\tilde H}_{nL}
=\omega {\bf 1}_{M}\otimes L_{3} + 
g_{1}\sum_{j=1}^{n}\sigma_{1}^{(j)}\otimes (L_{+}+L_{-}) + 
\frac{\Delta}{2}\sum_{j=1}^{n}\sigma_{3}^{(j)}\otimes {\bf 1}_{L} + 
g_{2}\sum_{j=1}^{n}\mbox{cos}(\omega_{j}t+\phi_{j})\sigma_{1}^{(j)}\otimes 
{\bf 1}_{L}, 
\end{equation}
where $M=2^{n}$ and $\sigma_{k}^{(j)}$ ($k=1,\ 3$) is 
\[
\sigma_{k}^{(j)}=1_{2}\otimes \cdots \otimes 1_{2}\otimes \sigma_{k}\otimes 
1_{2}\otimes \cdots \otimes 1_{2}\ (j-\mbox{position}). 
\]
See \cite{SAW} as an another model similar to this (its model assumes the RWA 
from the starting point). 

\par \noindent 
In the near future we will attempt an attack to this model. We would like to 
construct C-NOT operators for each pair of atoms (this is a very 
important subject in realistic Quantum Computation). 

By the way, according to increase of the number of atoms (we are expecting 
at least $n=100$ in the realistic quantum computation) we meet a very severe 
problem called Decoherence, see for example \cite{MFr3} and its references. 
We unfortunately don't know how to control this at the present. 

A generalization of the model to N--level system (see for example \cite{KF3}, 
\cite{KF6}, \cite{FHKW}, \cite{KuF}) is now under consideration and will be 
published in a separate paper\footnote{The author believes that it is 
important for us to consider the N--level system to prevent (or lessen) 
the decoherence problem}.

\vspace{10mm}
\noindent
{\it Acknowledgment.}
The author wishes to thank Marco Frasca for his crucial suggestions. 
He also wishes to thank the graduate students Kyoko Higashida, 
Ryosuke Kato and Yukako Wada for some help. 
%


\end{document}